\documentclass[11pt,twoside]{article}
\usepackage{./asp2014}
\usepackage{color}

\aspSuppressVolSlug
\resetcounters

\newcommand\degd{\ifmmode^{\circ}\!\!\!.\,\else$^{\circ}\!\!\!.\,$\fi}

\newcommand{\msun}{{\rm\ M_{\sun}}}

\newcommand{\ngvla}{ngVLA}

\begin{document}

\title{Science with Pulsar Timing Arrays and the \ngvla}
\author{The NANOGrav Collaboration
\affil{}
\email{info@nanograv.org}
}

\section{Summary}
Pulsar timing arrays (PTAs) can be used to detect and study gravitational waves in the nanohertz band (i.e., wavelengths of order light-years).  This requires  high-precision, decades-long data sets  from sensitive, instrumentally stable telescopes.  NANOGrav and its collaborators in the International Pulsar Timing Array consortium are on the verge of the first detection of the stochastic background produced by supermassive binary black holes, which form via the mergers of massive galaxies.  By providing Northern hemisphere sky coverage with exquisite sensitivity and higher frequency coverage compared to the SKA, a Next-Generation Very Large Array (ngVLA) will be a fundamental component in the next phase of nanohertz GW astrophysics, enabling detailed characterization of the stochastic background and the detection of individual sources contributing to the background, as well as detections of (or stringent constraints on) cosmic strings and other exotica.  Here we summarize the scientific goals of PTAs and the technical requirements for the ngVLA to play a significant role in the characterization of the nanohertz gravitational wave universe.

\section{Introduction: Science with Pulsar Timing Arrays}

The recent detections of binary black hole \citep{2016PhRvL.116f1102A}  and binary neutron star \citep{2017PhRvL.119p1101A} mergers by the LIGO-Virgo collaboration provide spectacular confirmation of the existence of gravitational waves (GWs) at frequencies of a few hundred hertz. These discoveries have measured the speed of gravity to phenomenal precision, determined  the origin of half of the elements of the periodic table, and probed relativistic explosions, as well as
changing our understanding of stellar evolution and binary interactions.

\begin{figure}[t!]
\includegraphics[width=0.95\textwidth]{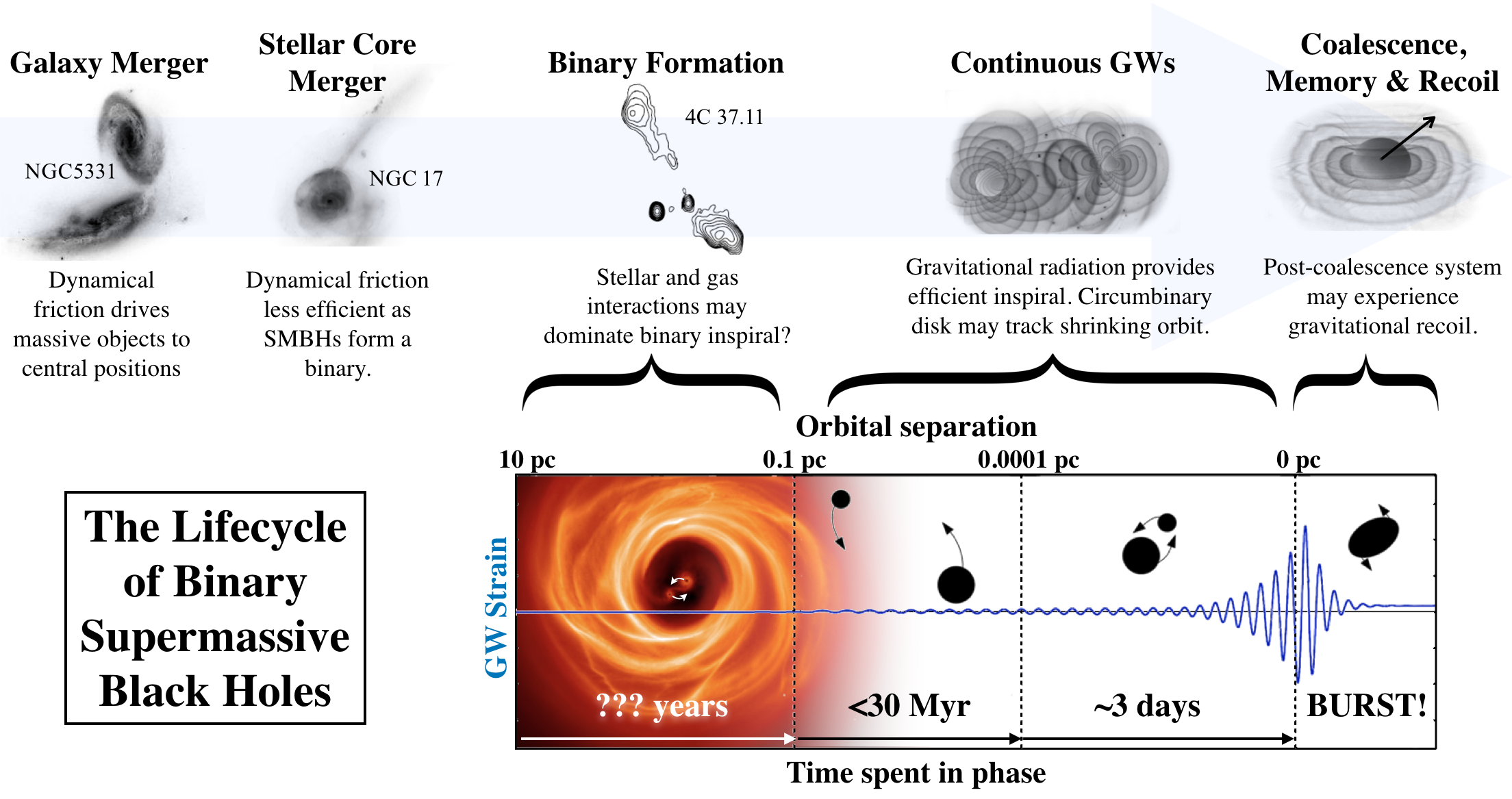}
\caption{Plot of the binary SMBH life-cycle, adapted from \citet{astro}. A significant unknown in binary evolution theory is the efficiency of inspiral from $\sim$10\,pc down to $\sim$0.1\,pc separations--the ``final parsec'' problem--after which the binary can coalesce efficiently due to gravitational wave emission.
Through sky imaging techniques, the ngVLA itself may discover widely separated binaries, and depending on its long-baseline sensitivity, could discover targets that are detectable by PTAs. Separately, the ngVLA could contribute to the sensitivity of PTAs, allowing the detection of GW emission from the same systems.
PTAs such as NANOGrav can detect supermassive ($>10^8\,\msun$) black hole binaries within $\sim$0.1\,pc separation (second panel in the lower figure). On rare occasion, PTAs may detect the permanent space-time deformation (GW memory) caused by a binary's coalescence \citep{favata09}.
Image credits: Galaxies, Hubble/STScI; 4C37.11, \citet{rodriguez}; Simulation visuals, C. Henze/NASA; Circumbinary accretion disk, C. Cuadra.}
\label{fig:SMBH}
\end{figure}

GWs at nanohertz frequencies---with wavelengths on the order of light-years---can be detected by a pulsar timing array (PTA), where a collection of stable millisecond pulsars (MSPs) is timed over a period of years to decades. These low-frequency GWs produce pulse arrival-time perturbations that are {\em spatially correlated} with a quadrupolar pattern on the sky \citep[e.g.,][]{hd83}, enabling the identification of a GW signal against a background of other effects including pulsar timing noise, stochastic pulse propagation in the interstellar and interplanetary media, and uncertainties in the position of the solar system barycenter. The most promising sources of GWs at nanohertz frequencies are supermassive black hole binaries (SMBHBs). Our current understanding of the formation of galaxies and the history of mass assembly in the universe requires the mergers of galaxies and, very probably, mergers of the supermassive black holes they host. Figure~\ref{fig:SMBH} shows the SMBHB life-cycle. A detection of the stochastic
background of GW emission produced by the mergers of SMBHBs is imminent.
The next step is studying individual binary systems through multimessenger probes.  By allowing the flexible allocation of collecting area to time pulsars of interest on a sustained and regular basis, the ngVLA will contribute significantly to the sensitivity of PTAs. Since all-sky coverage is essential, the ngVLA complements the Southern hemisphere coverage of the planned SKA. Further, as we show below, the optimal frequencies for timing distant, high dispersion measure pulsars are significantly higher than 1~GHz, making the ngVLA an ideal instrument for the expanding PTAs of the future.

In addition, through sky imaging techniques, the ngVLA may uncover widely separated SMBHBs, and depending on its long-baseline sensitivity, could discover targets that are simultaneously emitting gravitational waves in the PTA band. This would make the ngVLA a premier multi-messenger astronomy facility \textit{on its own}. Thus, it is possible that ngVLA could be a leading facility in the detection and study of binary SMBHs. It may contribute unique information to the full characterization of SMBH evolution that will be performed by current and future observatories such as LISA, LSST, ELT-class facilities, and the North American Nanohertz Observatory for Gravitational Waves (NANOGrav).

\section{Current State of the Art and Scientific Impact}

NANOGrav is a collaboration of astronomers using the world's most sensitive pulsar timing facilities (the Arecibo Observatory and the Green Bank Telescope) to monitor several tens of millisecond pulsars: we are expanding and improving a Galactic-scale detector for low-frequency GWs.
NANOGrav's recent 11-year data release~\citep{nano11} has allowed stringent limits to be placed on the stochastic background of GWs from SMBHB mergers \citep{nano11sb}, as well as meaningful constraints on more speculative source classes such as primordial GWs and cosmic strings that can form during phase transitions in the early Universe.

At present NANOGrav both competes and collaborates with its international partners, the European PTA and the Parkes PTA. As new telescopes (such as FAST and CHIME) come on line, we expect that collaboration under the umbrella of the International Pulsar Timing Array (IPTA) will play a more significant role. The leadership role of NANOGrav depends on continued access to telescope facilities for pulsar timing, along with new instrumentation that enables improvement in the sensitivity to gravitational waves.

In the next decade, we expect that the stochastic GW background will be successfully detected, either by NANOGrav or as an IPTA effort. The scientific focus will then shift to the precise spectral characterization of the stochastic GW background, measurement of its anisotropy, detection of continuous waves and bursts with memory from individual GW sources, joint observations of their electromagnetic counterparts, as well as constraints on the predictions from more exotic physics.   This will help address pressing questions in the joint evolution of supermassive black holes and their host galaxies (see \citealt{astro} for more details).

For instance, galaxy merger rates, SMBH-host co-evolution, dynamical relaxation timescales including the potential that SMBHBs may stall at wide separations, can all affect the amplitude scaling of the GW background.
The GW background spectrum might be detected with a shallow slope at frequencies $\lesssim$10\,nHz if SMBHBs have strong interactions with their environments (stars and gas) during the late stages of orbital evolution.  PTA constraints on or measurement of the background's amplitude and spectral shape can inform all of the above astrophysical uncertainties about the ensemble SMBHB population. A detection of the SMBHB background would provide the first comprehensive proof of the consensus view that SMBHs reside in most or all massive galaxies.
Moving past global characterizations, constraints on the background anisotropy may highlight actively interacting galaxy clusters or large-scale cosmic features.

Beyond the detection of the stochastic background, PTAs will reveal individually-resolvable continuous-wave sources as massive and nearby systems can be disentangled from the background. These detections will provide the most direct probe of the early-inspiral stage of a SMBHB, and can provide measurements of the binary's position, phase, and an entangled estimate of chirp mass and luminosity distance ($\mathcal{M}/D_{\rm L}$). If the host of a continuous-wave source can be identified with electromagnetic data, the mass/distance degeneracy can be broken by a redshift measurement; furthermore, this will enable a ``calibration'' of black hole mass/host galaxy relations at intermediate redshifts by the precise measurement of the central mass in these galaxies.
If detected, the pulsar term can permit temporal aperture synthesis, also allowing $\mathcal{M}$ and $D_{\rm L}$ to be disentangled. Evolution of the waveform over PTA experimental durations is unlikely for SMBHBs; however, this would also disentangle the $\mathcal{M}/D_{\rm L}$ term.

In our roadmap for the next 15 to 20 years, the ngVLA will provide Northern hemisphere coverage for PTAs with superb sensitivity and be a fundamental component of the multi-telescope program needed for nanohertz GW science that will be conducted by NANOGrav and the IPTA.

\begin{figure}
\includegraphics[height=8cm]{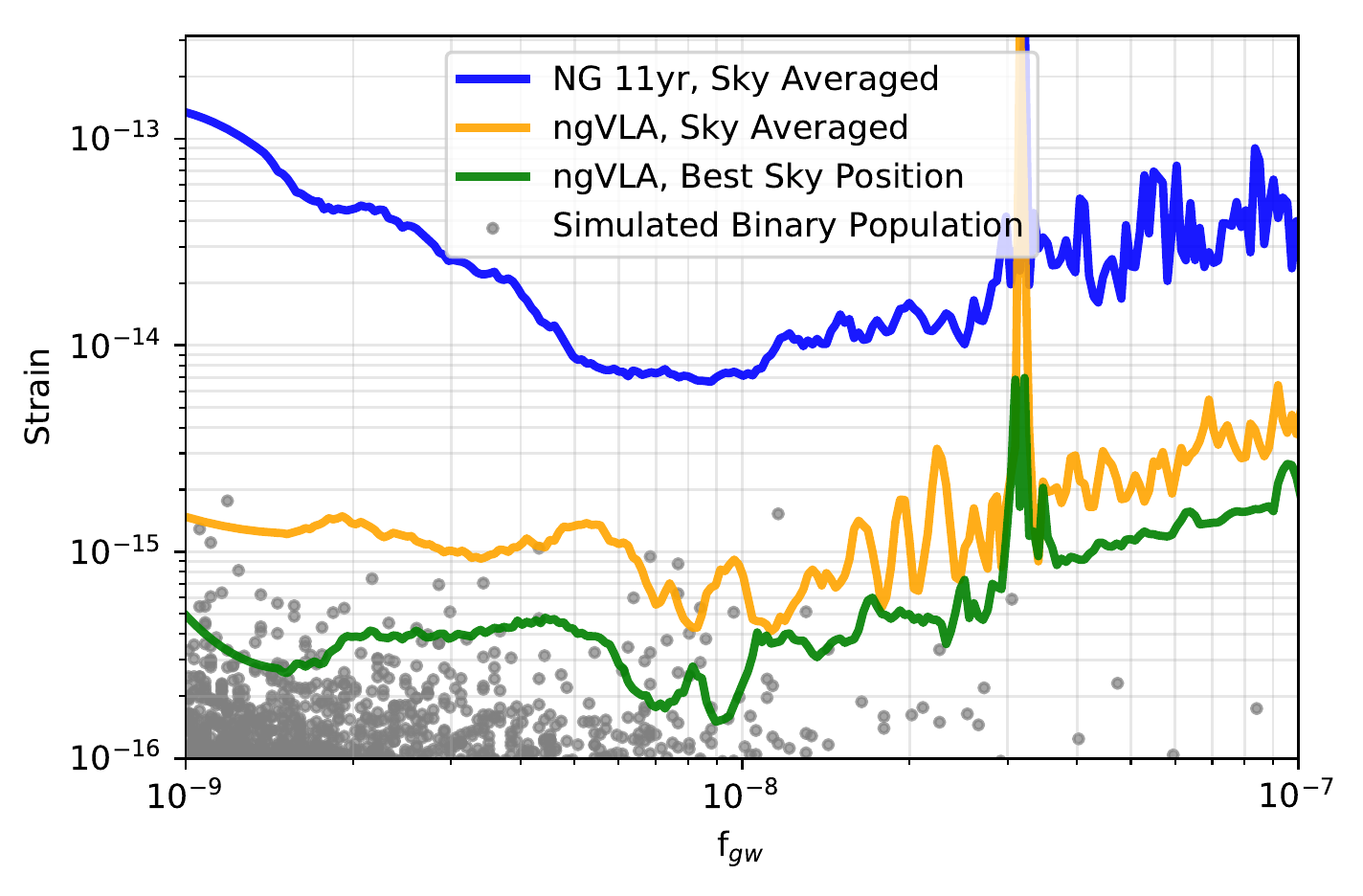}
\vspace{1mm}
\includegraphics[height=8cm]{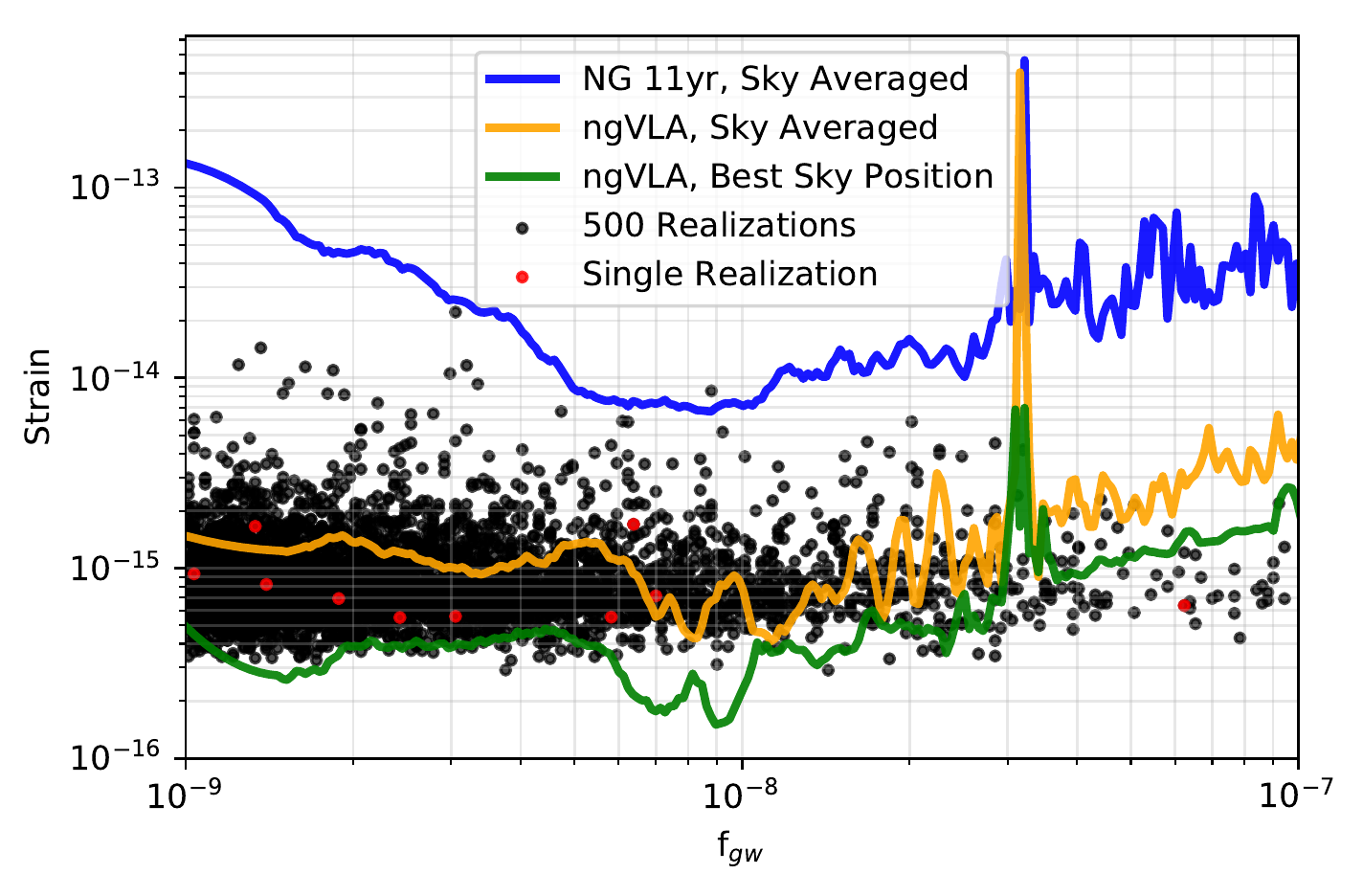}
\vspace{-3mm}
\caption{Plot of current sensitivity and sensitivity improvements for simulated observations using the ngVLA. In each panel, the top (blue) curve shows the current sky-averaged NANOGrav 11-yr 95\% upper limits. The bottom curves show the sky-averaged (orange) and best sky location (green) 95\% upper limits for simulated observations that assume a continuation of our current observing program followed by 10 years of ngVLA observations. {\bf Upper panel:} The gray points show a single simulation of the supermassive binary black hole population, the majority of which lie below our sensitivity limits. {\bf Lower panel:} The black points show the brightest 10 sources (only) in each of 500 simulations of the supermassive binary black hole population. Red points show the brightest 10 sources in one of these simulations.}
\label{fig:ngVLAUL}
\end{figure}

Figure~\ref{fig:ngVLAUL} shows our GW sensitivity goal with the ngVLA in terms of the gravitational wave
strain versus frequency. The top curve shows the sky-averaged 95\% upper limits for the NANOGrav 11-yr data set (our most current
data set). The bottom two curves (sky-averaged and best sky-location 95\% upper
limit curves) are computed from observation simulations that assume a
continuation of our current observing program followed by 10 years of ngVLA
observations. The black points show the brightest 10 sources in 500 simulations
of the SMBHB population in the universe. Depending on
the details of the simulation parameters, an average of 5-10 SMBHB lie above the upper limit curve for our best sky location using the
ngVLA. This estimate is also in broad agreement with \cite{mls+17}, who base their predictions on local massive galaxies.

\begin{figure}[t]
\hspace{-1cm}
\includegraphics[height=4.5cm]{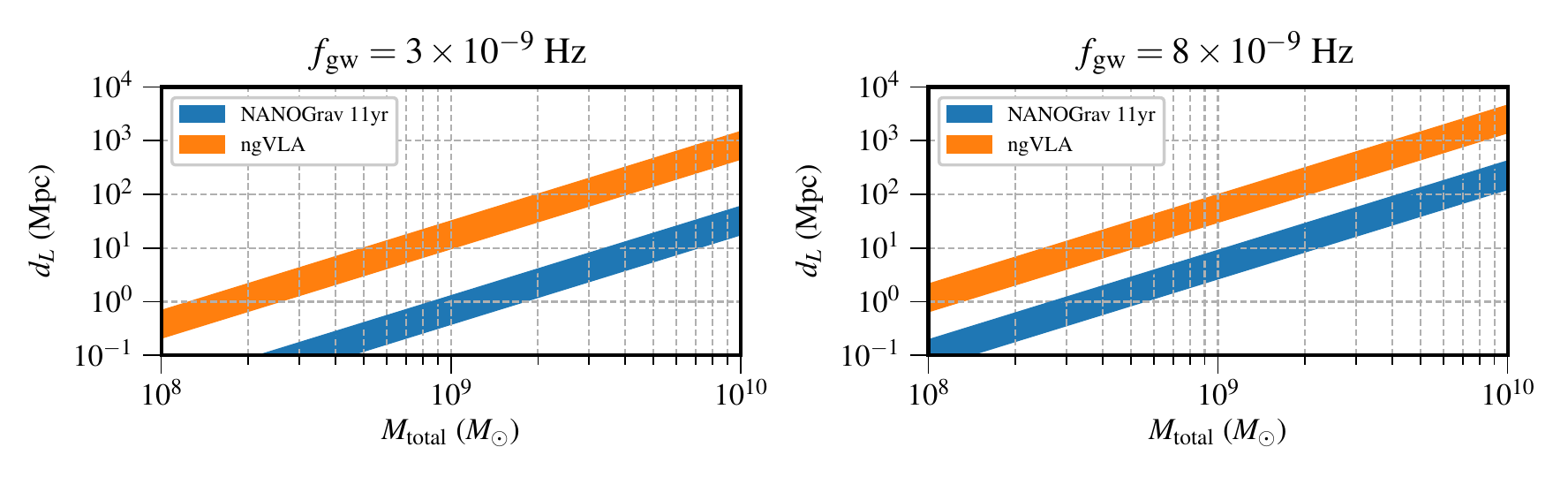}
\vspace{-8mm}
\caption{Plot of the luminosity distance versus the supermassive binary black hole total mass for our most current observations (the NANOGrav 11-yr data set, blue) and simulated ngVLA observations (orange). The luminosity distance (our astrophysical reach) increases by over an order of magnitude over all mass ranges.}
\label{fig:ngVLAdL}
\end{figure}

Figure~\ref{fig:ngVLAdL} shows the improvement in our detection range with the ngVLA in terms of the luminosity distance as a function of the supermassive black hole binary total
mass for two of our most sensitive frequencies. In the range of supermassive black hole binary masses relevant for pulsar timing arrays, the luminosity distance (our astrophysical reach) improves by over an order of magnitude. This increases the volume of the universe that is being probed for individual systems of nanohertz GW emission by a factor of over 1000.

Below we detail the improvements necessary to increase the sensitivity of our GW PTA detectors to achieve the scientific goals described here, and outline a future observational program. These improvements are then translated into a summary of the technical and usage requirements for the ngVLA to play a leading role in PTA science.

\section{Improving Nanohertz GW Detectors}

The key features of the ngVLA that are relevant to PTAs are its high sensitivity, Northern hemisphere sky-coverage, broad spectral coverage,
and the ability to multiplex its sensitivity into subarrays.  The last is a game-changing advantage over large single-dish telescopes.   We discuss two of the more important advances possible with the ngVLA below.

\subsection{Subarrays and the Fight Against Jitter}
\begin{figure}
\centering
\includegraphics[width=0.9\textwidth]{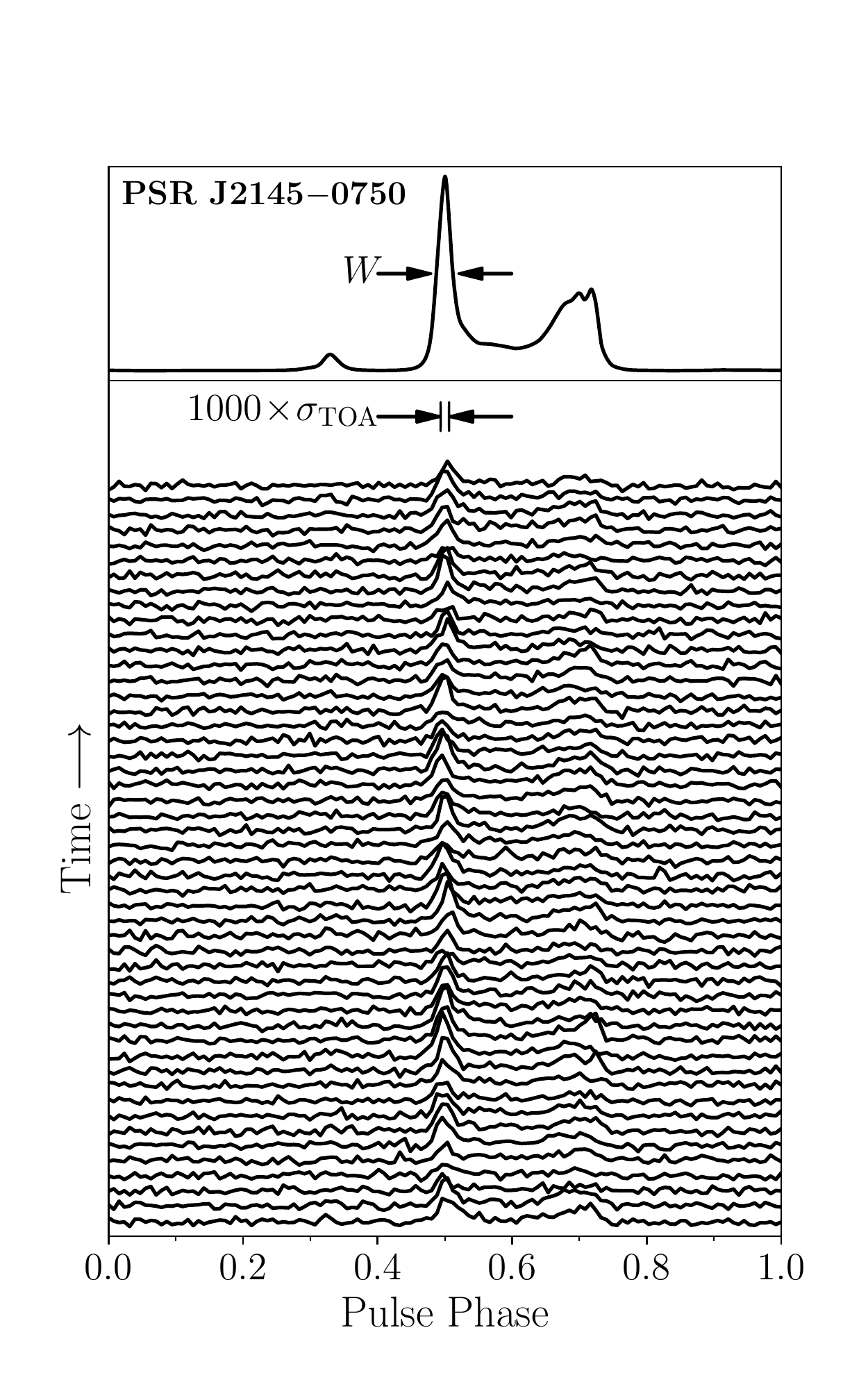}
\caption{Pulse profiles for PSR~J2145$-$0750, a 16.05~ms pulsar, at 820~MHz. The long-term average profile is shown at the top. The waterfall plot consists of a sequence of 10-pulse average profiles, illustrating the effect of pulse jitter, which reduces with the (number of pulses)$^{-1/2}$.
The attained TOA precision $\sigma_{\rm TOA} \sim 0.1~\mu$s, much smaller than the pulse width $W \sim 395~\mu$s. High-sensitivity, broad-bandwidth observations with the ngVLA can reduce the arrival-time uncertainties drastically for many of our pulsars, allowing for improved GW characterization in the future.}
\label{fig:waterfall}
\end{figure}

The subarray capability of the ngVLA is particularly important because, unlike traditional imaging observations, pulsar timing observations do not allow sensitivity to be traded against integration time without constraints.
Pulsar timing involves time-tagging of pulse profiles obtained by averaging large numbers ($N \sim 10^5$ to $10^6$) of individual pulses.  While such averages are highly stable, the amplitude and pulse-phase jitter of individual pulses causes small deviations of the average shape, which  translate into arrival time errors $\propto N^{-1/2}$, as illustrated in Figure~\ref{fig:waterfall}.

Pulse jitter therefore sets a floor on the minimum number of pulses that must be averaged to achieve a specified arrival-time precision for the millisecond pulsars whose arrival time errors are dominated by jitter (rather than signal-to-noise ratio considerations).
For those objects,  the required integration time does not decrease with increased telescope sensitivity.   For faint pulsars where radiometer noise currently dominates timing errors, increased sensitivity can improve their arrival time estimation precision and in some cases they will then be jitter dominated.   Therefore the full sensitivity of the ngVLA can be used to improve the timing errors of weaker MSPs, while brighter MSPs lend themselves to observations with subarrays large enough so that the arrival times are jitter dominated, thus optimizing pulsar timing throughput with the ngVLA.

\subsection{Frequency Coverage and the Fight Against the ISM}
The ionized interstellar medium causes chromatic perturbations of arrival times that must be removed  using multifrequency observations at frequencies that are optimized for each MSP.   Currently, there is not much flexibility in the choice of observing bands used for timing measurements, implying that there are significant opportunities for improving arrival times and the sensitivity to GWs with future instrumentation and at the ngVLA.

Pulsars are typically brighter at lower frequencies ($S_\nu \propto \nu^{\alpha}$; $\alpha \sim -1.6$) but considerations of receiver and sky background noise along with  interstellar effects   drive the range of observation frequencies (0.4 to 2~GHz) currently used by NANOGrav at Arecibo and the GBT.  %, as shown in Figure~\ref{fig:epochs}.
 Typically, a given MSP is observed at both a low frequency (typically $<1\,$GHz) and a high frequency ($>1$\,GHz) in order to fit for the dispersion measure (DM), which corresponds to the integrated line-of-sight electron column density (DM $\equiv \int_0^s n_e\,ds$). For the very high precision timing required for PTAs, the time variability of the DM requires contemporaneous measurements over a wide frequency range  at each observation epoch.

\begin{figure}[ht]
\includegraphics[width=0.49\textwidth]{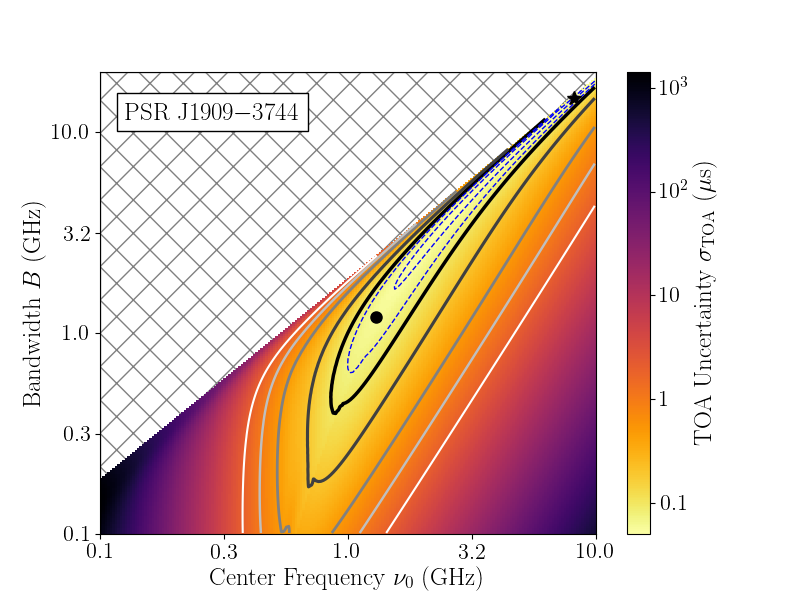}
\includegraphics[width=0.49\textwidth]{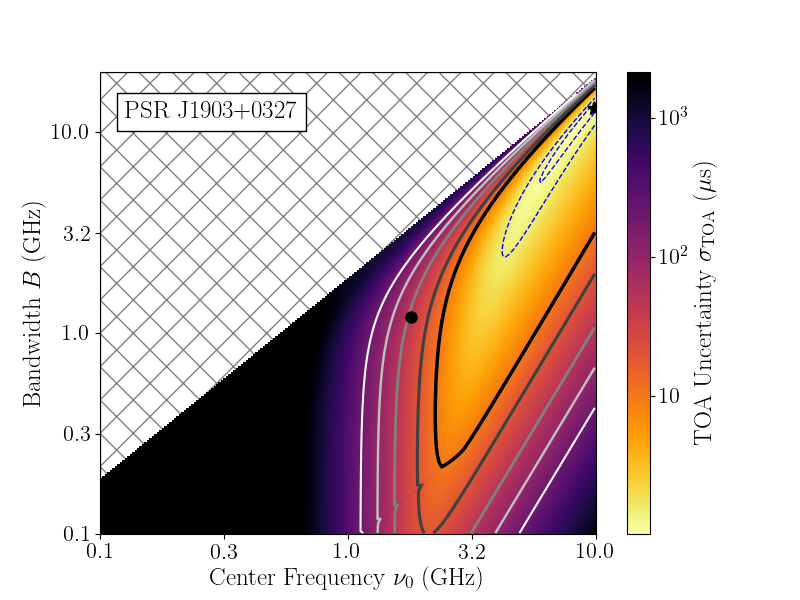}
\caption{The uncertainty in pulse time of arrival measurements as a function of observing center frequency and bandwidth for two millisecond pulsars currently being observed by NANOGrav, adapted from our work in \citet{lmc+18}. {\bf Left:} PSR~J1909--3744, which is one of the best-timed pulsars. Solid contours indicate TOA uncertainties of 2, 1, 0.5, 0.2, and 0.1~$\mathrm{\mu s}$, in order of increasing darkness and thickness. The minimum TOA uncertainty (black star) is $\sigma_{\rm TOA}(\nu_0=8.1~\mathrm{GHz},B=14.8~\mathrm{GHz}) = 50~\mathrm{ns}$ and the estimate given the current frequency coverage (black circle) is  $\sigma_{\rm TOA}(\nu_0=1.3~\mathrm{GHz},B=1.2~\mathrm{GHz}) = 59~\mathrm{ns}$. The two dashed blue contours represent a 10\% and 50\% increase above the minimum $\sigma_{\rm TOA}$. {\bf Right:} PSR~J1903+0327, which is the pulsar with the highest dispersion measure (297.52~pc~cm$^{-3}$) currently observed by NANOGrav. Contours indicate TOA uncertainties of 200, 100, 50, 20, and 10~$\mathrm{\mu s}$, in order of increasing darkness and thickness. The minimum TOA uncertainty (black star) is $\sigma_{\rm TOA}(\nu_0=9.8~\mathrm{GHz},B=13.2~\mathrm{GHz}) = 1.0~\mathrm{\mu s}$ and the estimate given the current frequency coverage (black circle) is  $\sigma_{\rm TOA}(\nu_0=1.8~\mathrm{GHz},B=1.2~\mathrm{GHz}) = 44.0~\mathrm{\mu s}$. Higher observing  frequencies with larger bandwidths potentially allow significant improvements in timing precision.}
\label{fig:toa}
\end{figure}

The optimization of  pulsar timing observations with respect to radio observing bands has been investigated in detail by \citet{lmc+18}, leading to the  conclusion that MSP timing can be improved significantly by using higher frequencies and larger bandwidths.
As shown in Figure~\ref{fig:toa}, higher frequencies and larger bandwidths can produce higher precision time-of-arrival (TOA) measurements than currently achieved. The effect is especially pronounced for pulsars at higher DMs, where propagation effects are increasingly important (Figure~\ref{fig:toa}, right panel). As the NANOGrav PTA expands, newer pulsar discoveries are likely to be fainter and more distant, so higher sensitivities and higher frequencies will become critical over time. However, access to frequencies down to 1--1.5~GHz remains essential; we note that with wide enough available bandwidths, dual-frequency observations may no longer be required for DM estimation, leading to large gains in efficiency.

These requirements can be addressed using  the ngVLA's  continuous frequency coverage in the overall $\sim 1$~to 10~GHz range of interest for PTAs.

\section{Observational Program}

The current NANOGrav timing program \citep{nano11} uses integration times of 0.25--0.5~hr per pulsar at Arecibo and Green Bank with cadences of weekly (for the best pulsars) to once every three weeks, and allocating similar integration times with a selected fraction of the ngVLA collecting area offers more flexibility and higher efficiency compared to steering the entire collecting area of a large single dish to a sequence of pulsar timing targets.

Thus we envisage two usage modes of the ngVLA for  PTA observations:  (1) A sole-user mode, where the array is used in full on a single object  or with up to $\sim 5$~subarrays, each observing a pulsar of interest; or (2) A shared-user mode, where a single phased subarray $\gtrsim$20\% of the ngVLA collecting area (i.e., comparable to the current GBT or more), depending on the pulsar flux density, is used to observe a single pulsar. As an aside, we note that sub-arraying reduces the net computational load, since baselines are not correlated across sub-arrays. Thus, as long as the correlator resources can be flexibly re-deployed, no extra computation resources will be required for shared-user or multi-target phased sub-array operation.

\smallskip\noindent
{\bf Frequency and Bandwidth:} As described above, frequency agility in the overall 1 to 10~GHz range is needed.

\smallskip\noindent
{\bf Observing Cadence:} NANOGrav currently observes $\sim$70 pulsars, with each pulsar in the array being observed (approximately) every three weeks. Beyond the detection of the stochastic GW background, the future PTA scientific program requires both an increasing number of pulsars in the array and an improved observing cadence.

A straw-man future observing program might involve $\sim$200 pulsars distributed over the sky. If each pulsar can be timed with a 20\% sub-array of the ngVLA for $\sim$0.5~hr every week, the total NANOGrav observing program may require $\sim$20~hr/week of the full array. However, such estimates require two important caveats. (1) The PTA observing program needs to be sustained for years---see, e.g., our recent data release spanning 11 years \citep{nano11}. (2) As the PTA expands, future pulsar additions to the array may be disproportionately fainter, requiring larger integration times. Instead, it appears more likely that NANOGrav will rely on the ngVLA to time the most critical and faintest pulsars, with the remainder being timed at partner facilities.

\smallskip\noindent
{\bf Pulsar Searching:} The PTA requirements for the ngVLA described here do {\em not} include the capability of blind, full field-of-view pulsar searches, which impose enormous computational loads. However, sensitivity to GWs increases with the number of pulsars, and requires a distribution of sources all over the sky for a good GW detector response function. It is likely that the discovery of new pulsars will be dominated by single dish telescopes and by hybrid methods, where multi-wavelength imaging is used to identify candidate compact sources that are then searched for pulsations. (A pulsar that is suitable for timing as part of a PTA will be easily identified as a compact source in radio sky survey images, or potentially via a high-energy counterpart.)

\smallskip\noindent
{\bf Complementarity with the SKA:} In contrast to the ngVLA, pulsar searching will be part of the core scientific mission of the planned SKA, and it will detect new pulsars to be timed as part of the future PTA. However, the varied scientific demands on the SKA imply that it is unlikely to accommodate regular pulsar timing observations at the required cadence for an all-sky PTA with a large number of pulsars. The ngVLA program described here will provide unique Northern hemisphere coverage, as well as the high frequencies needed to optimally observe distant, high-DM pulsars, as shown in Figure~\ref{fig:toa}. We expect to continue our current practice of cross-observing of the best few pulsars (e.g., PSR~J1909$-$3744 is currently observed by Australian, European, and North American telescopes) in order to address systematics and build a robust, fully-integrated international timing program.

\smallskip\noindent
{\bf Multi-messenger Coverage:} Elsewhere in this volume, chapters by Taylor and Simon (``From Megaparsecs To Milliparsecs: Galaxy Evolution and Supermassive Black Holes with NANOGrav and the ngVLA'') and Burke-Spolaor et al.\ (``Supermassive Black Hole Pairs and Binaries'') discuss the complementary capabilities of the ngVLA to not only contribute to the detection of GWs, but to observe the radio signatures of the selfsame targets. This capability potentially makes the ngVLA a comprehensive facility for multi-messenger studies of binary SMBHs by itself. In addition, the next generation of Extremely Large Telescopes and space-based optical/infrared and high energy surveys will reveal electromagnetic counterparts to the SMBHB population probed by PTA observations.

\section{Technical Requirements for PTA Observations with the ngVLA}

Given the more general programmatic issues described above, here we summarize the technical requirements for the ngVLA in order to enable PTA observations. We note that at present, none of these requirements represents a ``tall pole'' or driver for the ngVLA specifications.

\begin{description}

\item[Phased Sub-Arrays:] Multiple independent sub-arrays are required, up to $\sim$10, with each sub-array providing an independent phased array beam. It is preferable if each sub-array can maintain phase coherence over the duration of a pulsar timing observation, possibly using real-time self-calibration strategies with in-field calibrator sources.

\item[Frequency coverage and Bandwidth:] As described above and investigated further in \citet{lmc+18}, wide-band frequency coverage is required and coverage down to 1--1.5~GHz is essential. We note that some pulsars with large and time-variable DMs may be suited to  simultaneous dual-band observations using two independent sub-arrays, a capability of the ngVLA that is simply not available at any single-dish facility.

\item[Correlator and Computation needs:] Correlator specifications for pulsar observations are described in an ngVLA memo by Demorest et al., but in broad outline, phased array beams that can be sampled at 50~$\mu$s with 0.5~MHz channels are sufficient for PTA requirements. The output of the phased array beam will be coherently de-dispersed (i.e., a digital filter will be applied to remove the known average pulse dispersion as a function of frequency) in real time, before sampling. Given that cross-correlation for imaging will not be routinely required (and in any case, dishes observing different fields will never be correlated against each other), PTA observations using sub-arrays will be far less computationally challenging than full field-of-view, full-resolution imaging observations.

\item[Polarization Calibration:]  Emission from millisecond pulsars is typically polarized, with linear polarizations at the few--50\% level \citep[e.g.,][]{ymv+11}. The phased array beams will thus require polarization calibration (and more importantly, polarization stability) at the 5--10\% level, preferably better.

\item[Clock Stability:] The long-term clock stability requirement for pulsar timing observations is currently met by tying observatory masers to GPS time. On the short term, clock stability requirements for pulsar timing are exceeded by the requirements for high-frequency VLBI.

\item[Data Management and Curation:] PTA observations are a long-term enterprise, and the scientific value of the data set increases with time baseline as wider ranges of GW frequencies are probed to higher sensitivities. NRAO has an exemplary track record of incorporating infrastructure for data management over decade-long timescales at the VLA, but pulsar observations (e.g. at the GBT) have typically not been included in these plans. For the ngVLA, the volume of PTA observations is expected to be dwarfed by full-field visibility storage requirements, and a long term archival and curation plan is both essential and straightforward.

\end{description}

\section{Conclusions}

PTAs have the potential to open a new window on the gravitational wave spectrum, probing nanohertz emission from the supermassive binary black hole mergers that accompany mass assembly in the universe, as well as other, more exotic, sources. Due to its exquisite sensitivity and northern hemisphere sky-coverage, the ngVLA will play a key role in PTA observations as long as certain key requirements are met: most importantly, independently phased sub-arrays and frequency coverage down to 1--1.5~GHz. None of the requirements or constraints pose a significant obstacle given the existing specifications of the ngVLA.

\acknowledgments

The NANOGrav Physics Frontiers Center is supported by the National Science Foundation award number 1430284.

%\bibliography{myrefs}

\begin{thebibliography}{}
\expandafter\ifx\csname natexlab\endcsname\relax\def\natexlab#1{#1}\fi
\expandafter\ifx\csname url\endcsname\relax
  \def\url#1{\texttt{#1}}\fi
\expandafter\ifx\csname urlprefix\endcsname\relax\def\urlprefix{URL }\fi
\providecommand{\eprint}[2][]{\url{#2}}

\bibitem[{{Abbott} et~al.(2016){Abbott}, {Abbott}, {Abbott}, {Abernathy},
  {Acernese}, {Ackley}, {Adams}, {Adams}, {Addesso}, {Adhikari}
  et~al.}]{2016PhRvL.116f1102A}
{Abbott}, B.~P., {Abbott}, R., {Abbott}, T.~D., {Abernathy}, M.~R., {Acernese},
  F., {Ackley}, K., {Adams}, C., {Adams}, T., {Addesso}, P., {Adhikari}, R.~X.,
  et~al. 2016, Physical Review Letters, 116, 061102. \eprint{1602.03837}

\bibitem[{{Abbott} et~al.(2017){Abbott}, {Abbott}, {Abbott}, {Acernese},
  {Ackley}, {Adams}, {Adams}, {Addesso}, {Adhikari}, {Adya}
  et~al.}]{2017PhRvL.119p1101A}
{Abbott}, B.~P., {Abbott}, R., {Abbott}, T.~D., {Acernese}, F., {Ackley}, K.,
  {Adams}, C., {Adams}, T., {Addesso}, P., {Adhikari}, R.~X., {Adya}, V.~B.,
  et~al. 2017, Physical Review Letters, 119, 161101. \eprint{1710.05832}

\bibitem[{{Arzoumanian} et~al.(2018{\natexlab{a}}){Arzoumanian}, {Baker},
  {Brazier}, {Burke-Spolaor}, {Chamberlin}, {Chatterjee}, {Christy}, {Cordes},
  {Cornish}, {Crawford}, {Thankful Cromartie}, {Crowter}, {DeCesar},
  {Demorest}, {Dolch}, {Ellis}, {Ferdman}, {Ferrara}, {Folkner}, {Fonseca},
  {Garver-Daniels}, {Gentile}, {Haas}, {Hazboun}, {Huerta}, {Islo}, {Jones},
  {Jones}, {Kaplan}, {Kaspi}, {Lam}, {Lazio}, {Levin}, {Lommen}, {Lorimer},
  {Luo}, {Lynch}, {Madison}, {McLaughlin}, {McWilliams}, {Mingarelli}, {Ng},
  {Nice}, {Park}, {Pennucci}, {Pol}, {Ransom}, {Ray}, {Rasskazov}, {Siemens},
  {Simon}, {Spiewak}, {Stairs}, {Stinebring}, {Stovall}, {Swiggum}, {Taylor},
  {Vallisneri}, {van Haasteren}, {Vigeland}, {Zhu}, \& {The NANOGrav
  Collaboration}}]{nano11sb}
{Arzoumanian}, Z., {Baker}, P.~T., {Brazier}, A., {Burke-Spolaor}, S.,
  {Chamberlin}, S.~J., {Chatterjee}, S., {Christy}, B., {Cordes}, J.~M.,
  {Cornish}, N.~J., {Crawford}, F., {Thankful Cromartie}, H., {Crowter}, K.,
  {DeCesar}, M., {Demorest}, P.~B., {Dolch}, T., {Ellis}, J.~A., {Ferdman},
  R.~D., {Ferrara}, E., {Folkner}, W.~M., {Fonseca}, E., {Garver-Daniels}, N.,
  {Gentile}, P.~A., {Haas}, R., {Hazboun}, J.~S., {Huerta}, E.~A., {Islo}, K.,
  {Jones}, G., {Jones}, M.~L., {Kaplan}, D.~L., {Kaspi}, V.~M., {Lam}, M.~T.,
  {Lazio}, T.~J.~W., {Levin}, L., {Lommen}, A.~N., {Lorimer}, D.~R., {Luo}, J.,
  {Lynch}, R.~S., {Madison}, D.~R., {McLaughlin}, M.~A., {McWilliams}, S.~T.,
  {Mingarelli}, C.~M.~F., {Ng}, C., {Nice}, D.~J., {Park}, R.~S., {Pennucci},
  T.~T., {Pol}, N.~S., {Ransom}, S.~M., {Ray}, P.~S., {Rasskazov}, A.,
  {Siemens}, X., {Simon}, J., {Spiewak}, R., {Stairs}, I.~H., {Stinebring},
  D.~R., {Stovall}, K., {Swiggum}, J., {Taylor}, S.~R., {Vallisneri}, M., {van
  Haasteren}, R., {Vigeland}, S., {Zhu}, W.~W., \& {The NANOGrav Collaboration}
  2018{\natexlab{a}}, \apj, 859, 47. \eprint{1801.02617}

\bibitem[{{Arzoumanian} et~al.(2018{\natexlab{b}}){Arzoumanian}, {Brazier},
  {Burke-Spolaor}, {Chamberlin}, {Chatterjee}, {Christy}, {Cordes}, {Cornish},
  {Crawford}, {Thankful Cromartie}, {Crowter}, {DeCesar}, {Demorest}, {Dolch},
  {Ellis}, {Ferdman}, {Ferrara}, {Fonseca}, {Garver-Daniels}, {Gentile},
  {Halmrast}, {Huerta}, {Jenet}, {Jessup}, {Jones}, {Jones}, {Kaplan}, {Lam},
  {Lazio}, {Levin}, {Lommen}, {Lorimer}, {Luo}, {Lynch}, {Madison}, {Matthews},
  {McLaughlin}, {McWilliams}, {Mingarelli}, {Ng}, {Nice}, {Pennucci}, {Ransom},
  {Ray}, {Siemens}, {Simon}, {Spiewak}, {Stairs}, {Stinebring}, {Stovall},
  {Swiggum}, {Taylor}, {Vallisneri}, {van Haasteren}, {Vigeland}, {Zhu}, \&
  {The NANOGrav Collaboration}}]{nano11}
{Arzoumanian}, Z., {Brazier}, A., {Burke-Spolaor}, S., {Chamberlin}, S.,
  {Chatterjee}, S., {Christy}, B., {Cordes}, J.~M., {Cornish}, N.~J.,
  {Crawford}, F., {Thankful Cromartie}, H., {Crowter}, K., {DeCesar}, M.~E.,
  {Demorest}, P.~B., {Dolch}, T., {Ellis}, J.~A., {Ferdman}, R.~D., {Ferrara},
  E.~C., {Fonseca}, E., {Garver-Daniels}, N., {Gentile}, P.~A., {Halmrast}, D.,
  {Huerta}, E.~A., {Jenet}, F.~A., {Jessup}, C., {Jones}, G., {Jones}, M.~L.,
  {Kaplan}, D.~L., {Lam}, M.~T., {Lazio}, T.~J.~W., {Levin}, L., {Lommen}, A.,
  {Lorimer}, D.~R., {Luo}, J., {Lynch}, R.~S., {Madison}, D., {Matthews},
  A.~M., {McLaughlin}, M.~A., {McWilliams}, S.~T., {Mingarelli}, C., {Ng}, C.,
  {Nice}, D.~J., {Pennucci}, T.~T., {Ransom}, S.~M., {Ray}, P.~S., {Siemens},
  X., {Simon}, J., {Spiewak}, R., {Stairs}, I.~H., {Stinebring}, D.~R.,
  {Stovall}, K., {Swiggum}, J.~K., {Taylor}, S.~R., {Vallisneri}, M., {van
  Haasteren}, R., {Vigeland}, S.~J., {Zhu}, W., \& {The NANOGrav Collaboration}
  2018{\natexlab{b}}, \apjs, 235, 37. \eprint{1801.01837}

\bibitem[{{Favata}(2009)}]{favata09}
{Favata}, M. 2009, \apjl, 696, L159. \eprint{0902.3660}

\bibitem[{{Hellings} \& {Downs}(1983)}]{hd83}
{Hellings}, R.~W., \& {Downs}, G.~S. 1983, \apjl, 265, L39

\bibitem[{{Lam} et~al.(2017){Lam}, {McLaughlin}, {Cordes}, {Chatterjee}, \&
  {Lazio}}]{lmc+18}
{Lam}, M.~T., {McLaughlin}, M.~A., {Cordes}, J.~M., {Chatterjee}, S., \&
  {Lazio}, T.~J.~W. 2017, \apj, submitted. \eprint{1710.02272}

\bibitem[{{Mingarelli} et~al.(2017){Mingarelli}, {Lazio}, {Sesana}, {Greene},
  {Ellis}, {Ma}, {Croft}, {Burke-Spolaor}, \& {Taylor}}]{mls+17}
{Mingarelli}, C.~M.~F., {Lazio}, T.~J.~W., {Sesana}, A., {Greene}, J.~E.,
  {Ellis}, J.~A., {Ma}, C.-P., {Croft}, S., {Burke-Spolaor}, S., \& {Taylor},
  S.~R. 2017, Nature Astronomy, 1, 886. \eprint{1708.03491}

\bibitem[{{NANOGrav Astrophysics Working Group} et~al.(2018)}]{astro}
{NANOGrav Astrophysics Working Group}, et~al. 2018, in prep.

\bibitem[{{Rodriguez} et~al.(2006){Rodriguez}, {Taylor}, {Zavala}, {Peck},
  {Pollack}, \& {Romani}}]{rodriguez}
{Rodriguez}, C., {Taylor}, G.~B., {Zavala}, R.~T., {Peck}, A.~B., {Pollack},
  L.~K., \& {Romani}, R.~W. 2006, \apj, 646, 49. \eprint{astro-ph/0604042}

\bibitem[{{Yan} et~al.(2011){Yan}, {Manchester}, {van Straten}, {Reynolds},
  {Hobbs}, {Wang}, {Bailes}, {Bhat}, {Burke-Spolaor}, {Champion}, {Coles},
  {Hotan}, {Khoo}, {Oslowski}, {Sarkissian}, {Verbiest}, \& {Yardley}}]{ymv+11}
{Yan}, W.~M., {Manchester}, R.~N., {van Straten}, W., {Reynolds}, J.~E.,
  {Hobbs}, G., {Wang}, N., {Bailes}, M., {Bhat}, N.~D.~R., {Burke-Spolaor}, S.,
  {Champion}, D.~J., {Coles}, W.~A., {Hotan}, A.~W., {Khoo}, J., {Oslowski},
  S., {Sarkissian}, J.~M., {Verbiest}, J.~P.~W., \& {Yardley}, D.~R.~B. 2011,
  \mnras, 414, 2087. \eprint{1102.2274}

\end{thebibliography}

\end{document}